\documentstyle[twocolumn,graphicx,prb,aps]{revtex}
\input epsf
\tighten
\draft
\preprint{}
\begin{document}

\twocolumn[\hsize\textwidth\columnwidth\hsize \csname
@twocolumnfalse\endcsname 
\title{Models of Superconductivity in Sr$_2$RuO$_4$}

\author{Thomas~Dahm${}^a$, Hyekyung~Won${}^{b,c}$, Kazumi~Maki${}^c$}

\address{${}^a$Institut f\"ur Theoretische Physik, 
Universit\"at T\"ubingen, Auf der Morgenstelle 14, 
D-72076 T\"ubingen, Germany}
\address{${}^b$Department of Physics,
         Hallym University, 
         Chunchon 200-702, South Korea}
\address{${}^c$Department of Physics and Astronomy,
         University of Southern California, 
         Los Angeles, CA 90089-0484, U.S.A.}

\maketitle

\begin{abstract}
  Recent experimental data on purest Sr$_2$RuO$_4$ single
crystals clearly indicate the presence of nodes in the
superconducting order parameter. Here, we consider one
special $p$-wave order parameter symmetry and two
two-dimensional $f$-wave order parameter symmetries
having nodes within the RuO$_2$ plane. These states
reasonably describe both specific heat and
penetration depth data. We calculate the thermal conductivity
tensor for these three states and compare the results
with recent thermal conductivity data. This allows us
to single out one of these states being consistent with
both thermodynamic and thermal conductivity data:
the planar $f$-wave state having B$_{1g}\times$~E$_u$
symmetry.
\end{abstract}

]

\section{Introduction}
\label{secI}

The recently discovered superconductivity in Sr$_2$RuO$_4$ has been
interpreted in terms of a $p$-wave triplet superconducting state
having a full energy gap \cite{Maeno,Rice}. For example the
spontaneous spin polarization seen by muon spin rotation
experiments \cite{Luke} and the flat Knight-shift seen by
nuclear magnetic resonance (NMR) \cite{Ishida} are consistent
with spin triplet pairing. However, recent specific heat data
\cite{Nishizaki} and the superfluid density \cite{Bonalde} of
purest single crystals of Sr$_2$RuO$_4$ with $T_c \lesssim 1.5$K
clearly show low temperature behavior consistent with nodes in
the order parameter very similar to observations of $d$-wave
superconductivity in the high-$T_c$ cuprate superconductors 
\cite{Won,Hardy}.

Here we shall study three examples of two-dimensional (2D)
superconducting order parameters with spin triplet pairing
having nodes within the RuO$_2$ a-b plane. The first one is
the anisotropic $p$-wave state proposed by Miyake and Narikiyo
\cite{Miyake} with $\Delta(\vec{k}) \propto \sin(k_x a) \pm i 
\sin(k_y a)$. Here, $a$ denotes the lattice constant 
of the RuO$_2$ square lattice. In order to have a node with this state,
however, we have to stretch the Fermi wavevector $k_F$ towards 
the particular value of $k_F a = \pi$, while a more realistic 
value would be $k_F a = 2.7$ as judged from bandstructure 
calculations \cite{Mazin}. In the following we will denote this
particular $p$-wave state as the {\it nodal} $p$-wave state.
As the second and third example we
consider the planar $f$-wave states recently proposed by
Hasegawa et al \cite{Machida}. Here, the angular $\phi$ dependence
of the order parameter is given by $\Delta(\vec{k}) \propto
\cos(2\phi) e^{\pm i \phi}$ and $\Delta(\vec{k}) \propto
\sin(2\phi) e^{\pm i \phi}$, respectively.

\begin{figure}[thb]
  \begin{center}
    \includegraphics[width=0.75\columnwidth,angle=270]{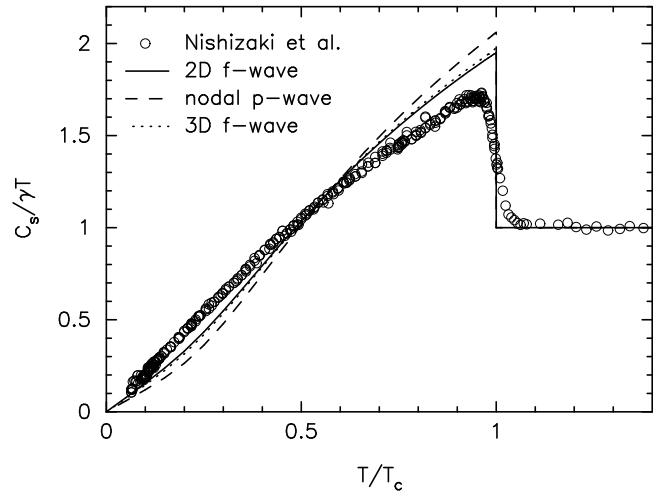}
    \vspace{.2cm}
    \caption{The specific heat $C_s/\gamma T$ as a function of
     $T/T_c$ for the 2D $f$-wave
     states (solid line) and the nodal $p$-wave state (dashed line) 
     considered in this work. Also shown are the experimental data
     by Nishizaki et al. \protect\cite{Nishizaki} (circles) and the 3D $f$-wave
     state considered in Ref. \protect\onlinecite{Yang} (dotted line).
     \label{Fig1} }
  \end{center}
\end{figure}

Within circular symmetric weak-coupling BCS theory one immediately 
realizes that
the thermodynamics of the latter two states is identical to
the one of $d$-wave superconductors \cite{Won}. We have worked
out the thermodynamics of the anisotropic, nodal $p$-wave state 
here as well.
In Figs. \ref{Fig1} and \ref{Fig2} we show our results for
the temperature dependence of the specific heat $C_s/\gamma T$ 
and the superfluid density $\rho_s(T)$ for the nodal $p$-wave
and the 2D $f$-wave states together with the experimental data.
For comparison, we also show the results of a 3D $f$-wave
state, considered by two of us recently \cite{Yang}. As is readily
seen, the 2D $f$-wave states give a better description
of the experimental data than the 3D $f$-wave or the nodal
$p$-wave states, though the differences between
the 2D $f$-wave and the 3D $f$-wave states are rather small.

Very recently the thermal conductivity of Sr$_2$RuO$_4$ in a
planar magnetic field has been studied \cite{Tanatar,Izawa}. Both
groups studied the thermal conductivity parallel to the a-axis
in a magnetic field within the a-b plane in a direction tilted
by an angle $\theta$ from the heat current. Both groups found
no appreciable angular dependence. This experimental result is
already inconsistent with the isotropic $p$-wave state having
a full enery gap and the 3D $f$-wave state state \cite{Yang}.
Indeed, we shall show that the thermal conductivity data is
consistent with only one of the three nodal states considered
here: the 2D $f$-wave state with angular dependence
$\cos(2\phi) e^{\pm i \phi}$. The two other states exhibit
rather large angular dependence and therefore are inconsistent
with the experiments.

In the next section we briefly summarize the thermodynamic
properties of the nodal $p$-wave superconductor with
$\Delta(\vec{k}) \propto \sin(k_x a) \pm i \sin(k_y a)$. In
many respects the results are very similar to the ones for
$d$-wave superconductors \cite{Won} and 3D $f$-wave
superconductors \cite{Yang}. Then we proceed to consider the
thermal conductivity in a planar magnetic field. The result
for the 2D $f$-wave state with angular dependence
$\cos(2\phi) e^{\pm i \phi}$ is very similar to the one in 
$d$-wave superconductors discussed recently in Ref. 
\onlinecite{Won2}.

\section{Thermodynamics of the nodal $p$-wave superconductor}
\label{secns}

We consider the superconducting order parameter given by
$\vec{\Delta}(\vec{k}) = \hat{d} \frac{\Delta}{s_M}
[ \sin(k_x a) \pm i \sin(k_y a) ]$ with $k_x a = \pi \cos(\phi)$
and $k_y a = \pi \sin(\phi)$ and the normalization
$s_M = \sqrt{2} \sin(\frac{\pi}{\sqrt{2}})=1.125$. This is
the model proposed in Ref. \onlinecite{Miyake} except that we 
have chosen the Fermi wavevector $k_F a = \pi$ in order to
have a node in $\vec{\Delta}(\vec{k})$. The quasi-particle Green
function in Nambu representation is given by

\begin{equation}
 G(k,\omega) = (i\omega - \xi_k \rho_3 - \Delta(k) \rho_1 \sigma_1)^{-1}
\end{equation}
where $\Delta(k)=\frac{\Delta}{s_M}[ \sin(k_x a) \pm i 
\sin(k_y a) ]$.

\begin{figure}[htb]
  \begin{center}
    \includegraphics[width=0.75\columnwidth,angle=270]{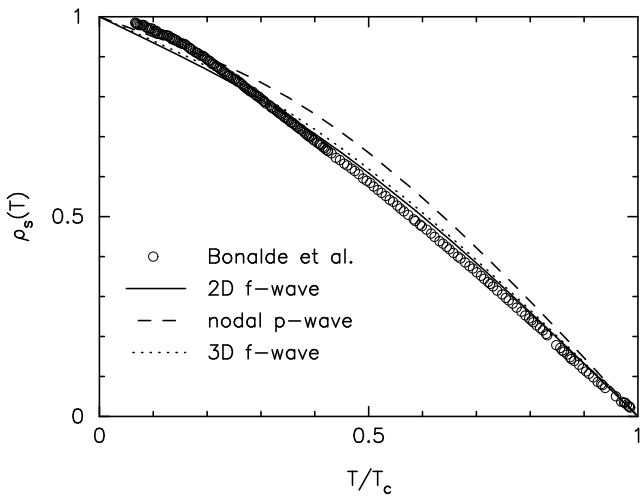}
    \vspace{.2cm}
    \caption{The superfluid density $\rho_s(T)$ as a function of
     $T/T_c$ for the 2D $f$-wave
     states (solid line) and the nodal $p$-wave state (dashed line) 
     together with the experimental data
     by Bonalde et al. \protect\cite{Bonalde} (circles) and the 3D $f$-wave
     state considered in Ref. \protect\onlinecite{Yang} (dotted line).
     \label{Fig2} }
  \end{center}
\end{figure} 

\begin{figure}[htb]
  \begin{center}
    \includegraphics[width=0.75\columnwidth,angle=270]{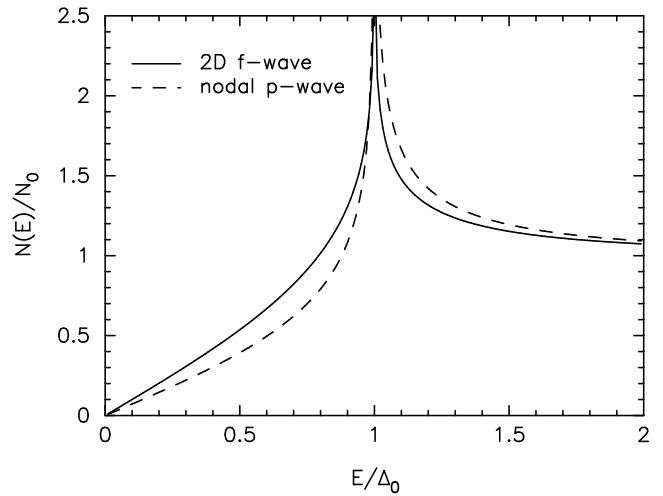}
     \vspace{.2cm}
    \caption{The density of states for the 2D $f$-wave state
     (solid line) and the nodal $p$-wave state (dashed line).
     \label{Fig3} }
  \end{center}
\end{figure}

Then the quasi-particle density of states is given by
\begin{eqnarray}
 N(E)/N_0 & = & {\rm Re} \langle \frac{E}{\sqrt{E^2-\Delta^2(k)}} 
\rangle \\ \nonumber
& = & \frac{4}{\pi} y \int_0^{\pi/4} d\phi \; {\rm Re}
\left( \frac{1}{\sqrt{y^2-f^2(\phi)}} \right)
\end{eqnarray}
where $f^2(\phi)=s_M^{-2} ( 1- \cos(\sqrt{2} \pi \cos\phi)
\cos(\sqrt{2} \pi \sin\phi))$, $y=E/\Delta$, and
$< >$ denotes an angular average. The
density of states is calculated and shown in Fig. \ref{Fig3}
together with the one for the 2D $f$-wave case. In
particular for $E/\Delta \ll 1$, the density of states
increases linearly as $N(E)/N_0 \simeq 0.7162 E/\Delta$,
while in the 2D $f$-wave case it varies like
$N(E)/N_0 \simeq E/\Delta$. Otherwise the two curves look
very similar.
Here, the gap equation
\begin{equation}
 \lambda^{-1} = \langle f^2 \rangle^{-1} \int_0^{E_c} dE
\langle \frac{f^2}{\sqrt{E^2-\Delta^2f^2(\phi)}}\rangle 
\tanh(\frac{E}{2T})
\end{equation}
has been solved numerically. In particular we find 
$\Delta(0)/T_c =2.00$,
which has to be compared with 2.14 in the 2D $f$-wave case.

The entropy $S$ is obtained from
\begin{eqnarray}
 S & = & -4 \int_0^\infty dE N(E) \left[ f \ln f + (1-f) 
\ln (1-f) \right]
\\ \nonumber
& = &  4 \int_0^\infty dE N(E) \left[ \beta E (1+e^{\beta E})^{-1}
+ \ln(1+e^{-\beta E}) \right]
\end{eqnarray}
with $f$ being the Fermi function and $\beta=1/T$.
Then 
the specific heat $C_s(T)$ is given by
\begin{equation}
  C_s(T) = T \frac{dS(T)}{ dT}
\end{equation}
$C_s(T)/\gamma T$ has been
shown in Fig. \ref{Fig1}. 

Finally, the superfluid density $\rho_s(T)$ is given by
\begin{equation}
 \rho_s(T) = 1- \frac{\beta \Delta}{2} \int_0^\infty dE \frac{N(E)}{N_0}
{\rm sech}^2 \left( \frac{\beta E}{2} \right)
\end{equation}
which behaves almost linearly in $T$ and is shown in Fig. \ref{Fig2}.
We note that at low temperatures an expansion of $\rho_s(T)$ leads
to $\rho_s(T)=1- 2 \ln 2 \times 0.7162 \frac{T}{\Delta} + \cdots$. 

\section{Thermal conductivity tensor in the a-b plane}
\label{secharm}

As shown in earlier experiments on YBCO, the thermal conductivity
tensor in a planar magnetic field is very sensitive to the
nodal directions \cite{Salamon,Aubin} and thus may be used to
further discriminate between the states studied above. We shall 
consider the thermal conductivity tensor in the vortex state
of the nodal $p$-wave and
the 2D $f$-wave separately. We will show that only one of these states
appears to be consistent with the angular independence observed
recently \cite{Tanatar,Izawa}.

\subsection{Nodal $p$-wave state}

The necessary theoretical scheme, neglecting vortex core scattering,
has been worked out during the
past few years \cite{Won2,Hirschfeld,Barash2}. We just apply this 
method for the present case. In particular for $\frac{H}{H_{c2}},
\frac{T^2}{\Delta^2} \ll 1$ in the superclean limit we obtain
\begin{eqnarray}
 \kappa_{xx}/\kappa_n & = & \frac{2}{\pi} \left( \frac{2 s_M}{\pi} 
\right)^2 \langle \left(1+\cos(2\phi) \right) x\rangle \langle x \rangle
\nonumber \\ 
& = &  \frac{2}{\pi} \left( \frac{2 s_M}{\pi} \right)^2  
\frac{v v' e H}{\Delta^2} F(\theta)
\end{eqnarray}
where $v$ and $v'$ are the Fermi velocities within the
a-b plane and perpendicular to it, respectively, and
$x=|{\bf v \cdot q}|/\Delta$ denotes the Doppler shift
due to the superflow around the vortex (see Ref.
\onlinecite{Won2}). $\kappa_n = \frac{\pi^2 T n}{6 \Gamma m}$ is the 
normal state thermal conductivity. The function $F(\theta)$ is given by
\begin{eqnarray}
 F(\theta) &=& \frac{2}{\pi^2} \sqrt{1+\sin^2 \theta} E\left( \frac{1}
{\sqrt{1+\sin^2 \theta}} \right) \times \nonumber \\
& &
\left( \sqrt{1+\sin^2 \theta} E\left( \frac{1}
{\sqrt{1+\sin^2 \theta}} \right) + \right. \nonumber \\
& &
\left. \sqrt{1+\cos^2 \theta} E\left( \frac{1}
{\sqrt{1+\cos^2 \theta}} \right) \right)
\label{EqnF}
\end{eqnarray}
with $E$ being the complete elliptic integral of the second kind and 
\begin{equation}
 \kappa_{xy}/\kappa_n = 0 \label{Eqkapp}
\end{equation}
In the present situation there will be no off-diagonal component,
because the heat current is parallel to the nodal direction.
The angular dependence of $\kappa_{xx}$ is given by the function
$F(\theta)$, which is shown in Fig. \ref{Fig6}. Surprisingly,
$\kappa_{xx}$ has a broad maximum for $\theta = \pi/2$. Also,
the anisotropy $\kappa_{xx}(\pi/2)/ \kappa_{xx}(0)=1.910$ is
quite strong. Therefore in view
of the thermal conductivity experimental data \cite{Tanatar,Izawa},
we have to reject this possibility.

\subsection{2D $f$-wave state}

\begin{figure}[thb]
  \begin{center}
    \includegraphics[width=0.75\columnwidth,angle=270]{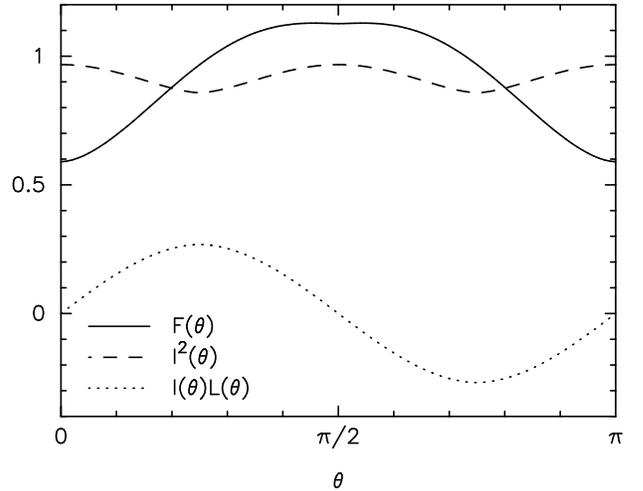}
    \vspace{.2cm}
    \caption{Angular variation of the functions $F(\theta)$, 
     $I^2(\theta)$, and $I(\theta)L(\theta)$ as defined in Eqs.
     (\protect\ref{EqnF}), (\protect\ref{EqnI}), and (\protect\ref{EqnL}). 
     The angular variation
     of the thermal conductivity $\kappa_{xx}$ for the nodal $p$-wave 
     state, as given by the function $F(\theta)$, is much stronger
     than for the $\cos(2\phi) e^{\pm i \phi}$ $f$-wave state 
     ($I^2(\theta)$) due to
     the different position of the nodes in the gap function.
     \label{Fig6} }
  \end{center}
\end{figure}

As already mentioned we consider the two states $\sin(2\phi) 
e^{\pm i \phi}$ and $\cos(2\phi) e^{\pm i \phi}$. The order
parameter $\propto \sin(2\phi) e^{\pm i \phi}$ has the same
nodal structure as the nodal $p$-wave state studied in the
last subsection and has been studied recently by Graf and 
Balatsky \cite{Graf}. Following the same procedure as above we 
find for the state $\sin(2\phi) e^{\pm i \phi}$
\begin{equation}
 \kappa_{xx}/\kappa_n  =  \frac{2}{\pi}
 \langle \left(1+\cos(2\phi) \right) x\rangle \langle x \rangle
 =   \frac{2}{\pi} 
\frac{v v' e H}{\Delta^2} F(\theta)
\end{equation}
and 
\begin{equation}
 \kappa_{xy} = 0 \label{Eqkapf1}
\end{equation}
where $F(\theta)$ has been shown in Fig. \ref{Fig6}.

Therefore, also the state $\sin(2\phi) e^{\pm i \phi}$
gives a rather large $\theta$ dependence, which is inconsistent
with the existent experiments \cite{Tanatar,Izawa}.

Finally, let us consider the state $\cos(2\phi) e^{\pm i \phi}$,
which has its nodes along the zone diagonal. As already noted,
this state has the same thermodynamics as a $d$-wave superconductor.
Further, the thermal conductivity tensor is now given by
\cite{Won2}
\begin{equation}
 \kappa_{xx}/\kappa_n = \frac{2}{\pi} 
\frac{v v' e H}{\Delta^2} I^2(\theta)
\end{equation}
and
\begin{equation}
 \kappa_{xy}/\kappa_n = -\frac{2}{\pi} 
\frac{v v' e H}{\Delta^2} I(\theta) L(\theta)
\label{Eqkapf2}
\end{equation}
where
\begin{eqnarray}
 I(\theta) &=& \frac{1}{\pi}
\left( \sqrt{\frac{3+s}{2}} E\left( \sqrt{\frac{2}
{3+s}} \right) + \right. \nonumber \\ & &
\left. \sqrt{\frac{3-s}{2}} E\left( \sqrt{\frac{2}
{3-s}} \right) \right)
\label{EqnI}
\end{eqnarray}
and 
\begin{eqnarray}
 L(\theta) &=& \frac{1}{\pi}
\left( \sqrt{\frac{3+s}{2}} E\left( \sqrt{\frac{2}
{3+s}} \right) - \right. \nonumber \\ & &
\left. \sqrt{\frac{3-s}{2}} E\left( \sqrt{\frac{2}
{3-s}} \right) \right)
\label{EqnL}
\end{eqnarray}
with $s=\sin(2\theta)$.

In Fig. \ref{Fig6} the angular 
dependences of the functions $(I(\theta))^2$ and
$I(\theta) L(\theta)$ are shown together with $F(\theta)$.
Thus, as in $d$-wave superconductors, this state exhibits a
fourfold symmetry in $\kappa_{xx}$. But the angular dependence
is about 10\% and may be compatible with the experiments
\cite{Tanatar,Izawa}. Thus, we conclude that the
$\cos(2\phi) e^{\pm i\phi}$ 2D $f$-wave state is the best
candidate to describe all of these experimental observations.

As mentioned above, our analysis of the thermal conductivity tensor
neglects vortex core scattering. At least in the high-$T_c$ compounds
in a small magnetic field and at low temperatures this contribution
can be neglected \cite{Hirschfeld,Chiao}. In Sr$_2$RuO$_4$ the
vortex core size is larger and at present it is unclear to what
extend this contribution plays a role. We expect that the
angular dependences shown in Fig. \ref{Fig6} will be weakened
both by vortex core scattering and finite temperatures, which will
improve agreement with the experiments.

As an additional check on the position of the nodes of the
order parameter we propose a measurement of the transverse
thermal conductivity $\kappa_{xy}$. As Eqs. (\ref{Eqkapp}) and
(\ref{Eqkapf1}) show, $\kappa_{xy}$ vanishes, if the nodes lie
along the a or b directions. However, we expect a finite
transverse thermal conductivity $\kappa_{xy}$ showing a
$\sin(2\theta)$ variation for the $\cos(2\phi) e^{\pm i\phi}$ 
$f$-wave state having its nodes along the zone diagonal,
as Eq. (\ref{Eqkapf2}) shows.

\section{Conclusions}
\label{seccon}

We compared one $p$-wave and two 2D $f$-wave superconducting states
with recent experimental data from purest crystals of Sr$_2$RuO$_4$.
We find that within weak-coupling theory the two 2D $f$-wave states 
give the closest description of the thermodynamic data.
Among these, the angular dependence of magnetotransport favors 
the $\cos(2\phi) e^{\pm i\phi}$ $f$-wave state, since the other
two states exhibit much stronger anisotropy than observed
experimentally. Therefore, among the simplest states 
the $\cos(2\phi) e^{\pm i\phi}$ $f$-wave state, having
B$_{1g}\times$~E$_u$ symmetry appears to be the best candidate for
superconductivity in Sr$_2$RuO$_4$.

\acknowledgments

We thank Y.~Maeno and I.~Bonalde for providing us with the digital
form of their experimental data, which were used in Fig. \ref{Fig1} 
and Fig. \ref{Fig2}.

We also thank K.~Izawa, Y.~Maeno, Y.~Matsuda, and M.~A.~Tanatar
for fruitful discussions on their ongoing experiments.
One of us (KM) thanks for the hospitality of N.~Schopohl and
the University of T\"ubingen where part of this work has been
done. HW acknowledges support from the Korean Science and
Engineering Foundation (KOSEF) through grant No. 1999-2-114-005-5.

\end{document}